\documentclass[pre,twocolumn,superscriptaddress,showpacs]{revtex4}
\usepackage{amsmath}
\usepackage{latexsym}
\usepackage{graphicx}

\begin{document}

\title{Comments on Sweeny and Gliozzi dynamics for simulations of Potts
models in the Fortuin-Kasteleyn representation}
\author{Jian-Sheng Wang}
\affiliation{Singapore-MIT Alliance and Department of Computational Science, 
National University of Singapore, Singapore 119260, Republic of Singapore}
\author{Oner Kozan}
\author{Robert H. Swendsen}
\affiliation{Department of Physics, Carnegie Mellon University, Pittsburgh, PA 15213}
\date{24 June 2002}

\begin{abstract}
We compare the correlation times of the Sweeny and Gliozzi dynamics for
two-dimensional Ising and three-state Potts models, and the
three-dimensional Ising model for the simulations in the percolation
representation. The results are also compared with Swendsen-Wang and Wolff
cluster dynamics. It is found that Sweeny and Gliozzi dynamics have
essentially the same dynamical critical behavior. Contrary to Gliozzi's
claim (cond-mat/0201285), the Gliozzi dynamics has critical slowing down
comparable to that of other cluster methods. For the two-dimensional Ising
model, both Sweeny and Gliozzi dynamics give good fits to logarithmic size
dependences; for two-dimensional three-state Potts model, their dynamical
critical exponent $z$ is $0.49 \pm 0.01$; the three-dimensional Ising model
has $z = 0.37 \pm 0.02$.
\end{abstract}

\pacs{05.50.+q, 05.10.Ln, 75.10.Hk}
\maketitle

\bigskip 


Cluster algorithms have played an interesting role in statistical physics,
both for their importance in constructing efficient computational algorithms
and their unusual dynamical properties. Very recently, Gliozzi has
introduced a new cluster algorithm, which he claims to be free of critical
slowing down \cite{gliozzi} . \ Since an understanding of the dynamics at the
critical temperature is central to both dynamic universality classes and
computational efficiency, we have made a comparison of the four main cluster
methods for Potts models due to Sweeny \cite{sweeny}, Swendsen and Wang 
\cite{swendsen-wang}, Wolff \cite{wolff}, and Gliozzi \cite{gliozzi}.

In 1983 Sweeny \cite{sweeny} simulated the Potts model \cite{wu} directly in the
Fortuin-Kasteleyn representation \cite{FK}, given by the probability
distribution of percolation bond configuration $\Gamma $, 
\begin{equation}
P(\Gamma )\propto \left( {\frac{p}{1-p}}\right) ^{b}q^{N_{c}},  \label{prob}
\end{equation}%
where $p=1-\exp \bigl(-J/(kT)\bigr)$, $J$ is coupling constant in the Potts
model, $k$ is Boltzmann constant, and $T$ is temperature; $b$ is the number
of bonds present on the (hypercubic) lattice, and $N_{c}$ is the number of
clusters of the percolation configuration.

\begin{figure}[tb]
\includegraphics[width=0.7\columnwidth]{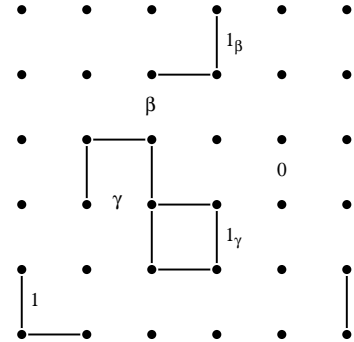}
\caption[bonds]{Different types of links: $1$ (occupied bond),
$0$ (absence of a bond), $\beta$ (unoccupied link that bridges two clusters),
$\gamma$ (unoccupied link that spans the same cluster), 
$1_\beta$ (bond, removing of which breaks the cluster), 
$1_\gamma$ (bond, removing of which does not break the cluster).}
\label{fig-0}
\end{figure}

Sweeny used a heat-bath rate of ``flipping'' the links from occupied to
empty or vice versa. Consider a particular link. We define the state $1$ to
mean a presence of a bond, $0$ for absence of a bond, $\beta $ for
unoccupied link such that the two nearest neighbor sites are on different
clusters, and $\gamma $ for unoccupied link that spans the same cluster,
see Fig.~\ref{fig-0}. In
addition, we define $1_{\beta }$ to be an occupied bond, the removal of
which leads to $\beta $, and similarly $1_{\gamma }$ is an occupied bond,
the removal of which leads to $\gamma $. \ \ Finally, a dot $\cdot $
represents an arbitrary state (0 or 1). Assuming a link is picked up at 
random, the transition rate for Sweeny's dynamics is 
\begin{subequations}
\begin{eqnarray}
w(\cdot \rightarrow 1_{\gamma })=p, &&\!\!\!\!w(\cdot \rightarrow \gamma
)=1-p, \\
w(\cdot \!\rightarrow \!1_{\beta })={\frac{p}{(1\!-\!p)q\!+\!p}},
&&\!\!\!\!w(\cdot \!\rightarrow \!\beta )={\frac{(1\!-\!p)q}{(1\!-\!p)q\!+\!p%
}}.
\end{eqnarray}%
Note that the transition probabilities do not depend on the initial state,
as it is generally the case for heat-bath algorithms.

Gliozzi also worked in the Fortuin-Kasteleyn representation \cite{FK}  and
proposed a variation on the flip rate \cite{gliozzi} that is only slightly
different from Sweeny's: 
\end{subequations}
\begin{subequations}
\begin{eqnarray}
w(1\rightarrow 1)=p, &&w(1\rightarrow 0)=1-p, \\
w(\gamma \rightarrow 1)=p, &&w(\gamma \rightarrow 0)=1-p, \\
w(\beta \rightarrow 1)=p/q, &&w(\beta \rightarrow 0)=1-p/q.
\end{eqnarray}

One can easily show that both of algorithms satisfy detailed balance with
respect to the equilibrium distribution, Eq.~(\ref{prob}). Both Sweeny and
Gliozzi's rates are applicable to model with real positive $q$ (not just
integer values), although Gliozzi's rate has the restriction that  $p/q \leq 1$.
We also note that Chayes and Machta proposed a cluster algorithm for 
noninteger $q$ that takes $O(L^d)$ operations per sweep
\cite{chayes}.

A key implementation issue in both algorithms is the determination of
whether two sites belong to the same cluster. For two dimensions, Sweeny
devised a very efficient method based on the special topology. In three and
higher dimensions, one has to contend with the shortcomings that a move
takes an amount of CPU time proportional to the typical size of the clusters.

To separate the issues of the intrinsic dynamics of the transition rates
from the efficiency of the numerical implementation, we present results for
the correlation times of both Sweeny and Gliozzi dynamics, measured in Monte
Carlo sweeps regardless of the actual computer time for a particular
implementation. We also compare them in the same way with Swendsen-Wang 
\cite{swendsen-wang} and Wolff \cite{wolff} cluster dynamics.

We consider the integrated correlation time, which is directly relevant to the
magnitude of statistical errors \cite{MK-Binder}. Instead of the usual
method of computing the autocorrelation function for, say, energy, 
\end{subequations}
\begin{equation}
f(t)={\frac{\langle E(0)E(t)\rangle -\langle E(0)\rangle ^{2}}{\langle
E(0)^{2}\rangle -\langle E(0)\rangle ^{2}}},
\end{equation}%
a totally equivalent way is to compute the variance of $M$ consecutive
terms of the sum of energies, 
\begin{equation}
E_{M}={\frac{1}{M}}\sum_{t=1}^{M}E(t).
\end{equation}%
We have for the variance \cite{landau-binder} 
\begin{equation}
\mathrm{var}(E_{M})={\frac{\tau _{M}\,\mathrm{var}(E_{1})}{M}}.
\end{equation}%
A straightforward variance estimator, i.e., the squared mean minus the
sample mean squared, is used. The usual correlation time $\tau $ is the
limit of $M$ going to infinity, and is related to the correlation function
by 
\begin{equation}
\tau =\lim_{M\rightarrow \infty }\tau _{M}=\sum_{t=-\infty }^{\infty }f(t).
\end{equation}%
Kikuchi et al \cite{okabe} have used a similar method to extract exponential
correlation times. Other conventions for specifying the correlation time are
in use. For example, M\"{u}ller-Krumbhaar and Binder \cite{MK-Binder} call 
$\tau ^{\prime }$ the correlation time, where  $\tau =1+2\tau ^{\prime }$,
and Baillie and Coddington \cite{baillie} use the definition $\tau ^{\prime
\prime }=\tau /2$.

\begin{figure}[tb]
\includegraphics[width=\columnwidth]{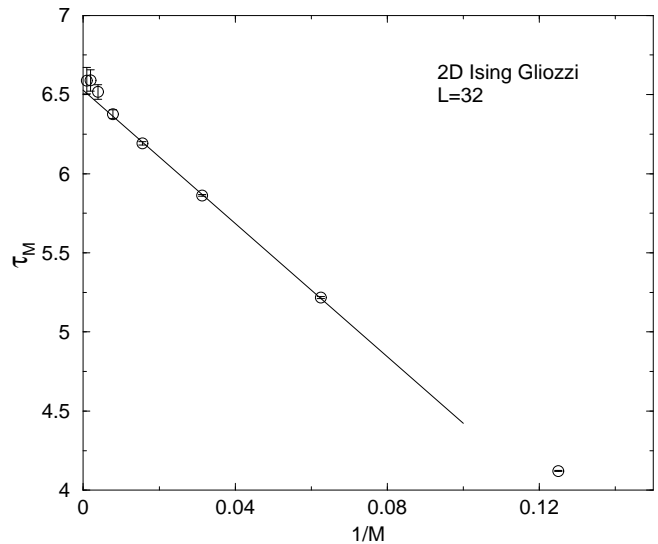}
\caption[tau]{$\protect\tau_M$ versus $1/M$ for the two-dimensional Ising model
with Gliozzi dynamics on a $32 \times 32$ lattice, with $8 \times 10^6$
Monte Carlo sweeps. The straight is an error weighted least-square fit. The
intercept gives the correlation time $\protect\tau = 6.53 \pm 0.01$.}
\label{fig-1}
\end{figure}

The truncated sum converges to its limit according to $1/M$. Thus we obtain
the limit with a plot of $\tau _{M}$ versus $1/M$, as shown in Figure~\ref%
{fig-1}. The method appears excellent for dynamics with relatively small
values of $\tau $. When $\tau $ is large, we have to go to much bigger
values of $M$, where the noise can dominate the signal, rendering the method
less useful.

\begin{table}
\caption{\label{table1}Correlation time $\tau$ for Sweeny and Gliozzi
dynamics at various system sizes.  The number in parentheses is standard 
error at the last digits.}
\begin{ruledtabular}
\begin{tabular}{lll}
$L$ &  Sweeny rate & Gliozzi rate\\
\hline
2D Ising\\
\hline
 2 & 2.6121(3)  &  2.9414(4) \\
 4 & 3.3280(3)  &  3.8722(7) \\
 8 & 4.006(2)   &  4.738(1)  \\
16 & 4.688(5)   &  5.607(5) \\
32 & 5.46(2)    &  6.51(3)  \\
64 & 6.21(3)    &  7.54(5)  \\
\hline
2D $q=3$ Potts\\
\hline
 2 & 3.0736(2)  & 3.8111(4) \\
 4 & 4.5375(8)  & 5.9245(5) \\
 8 & 6.424(2)   & 8.604(3) \\
16 & 9.06(2)    & 12.18(1) \\
32 & 12.56(5)   & 17.10(5) \\
64 & 17.5(3)    & 24.0(2) \\
\hline
3D Ising\\
\hline
2  & 3.0242(3)& 3.3456(4)  \\
4  & 4.254(2) & 4.814(16) \\
8  & 5.594(8) & 6.375(8)  \\
16 & 7.1(1)   & 8.0(2)   \\
32 & 9.0(5)   & 11.4(10) \\
\end{tabular}
\end{ruledtabular}
\end{table}

\begin{figure}[t]
\includegraphics[width=\columnwidth]{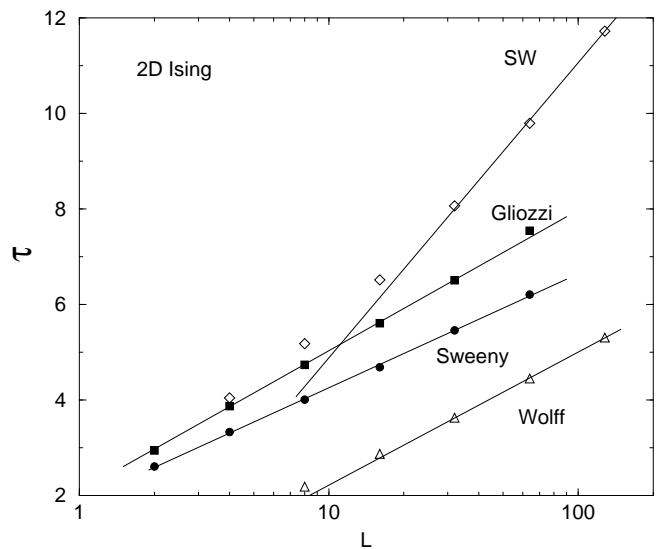}
\caption[2D-Ising]{Correlation times $\protect\tau$ versus lattice linear
dimension $L$ on a semi-logarithmic scale for the two-dimensional Ising
model at $T_c$. The circles are for Sweeny, squares for Gliozzi, diamonds
for Swendsen-Wang, and triangles for Wolff dynamics.  Straight lines are fits
to asymptotic slopes at large $L$.  The Swendsen-Wang data are
from ref.~\protect\cite{WangMCQMC} and Wolff data are from
ref.~\protect\cite{baillie}.}
\label{fig-2}
\end{figure}

\begin{figure}[tbp]
\includegraphics[width=\columnwidth]{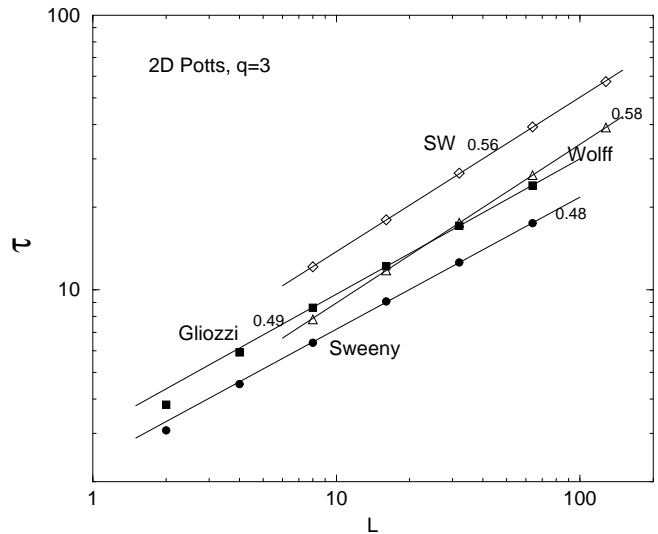}
\caption{Same as Fig.~\ref{fig-2}, but on a double logarithmic scale for the
two-dimensional three-state Potts model at $T_c$. The number on the line
indicates the slope of the line.  The Swendsen-Wang and Wolff data are from
ref.~\protect\cite{baillie}.}
\label{fig-3}
\end{figure}

\begin{figure}[tbp]
\includegraphics[width=\columnwidth]{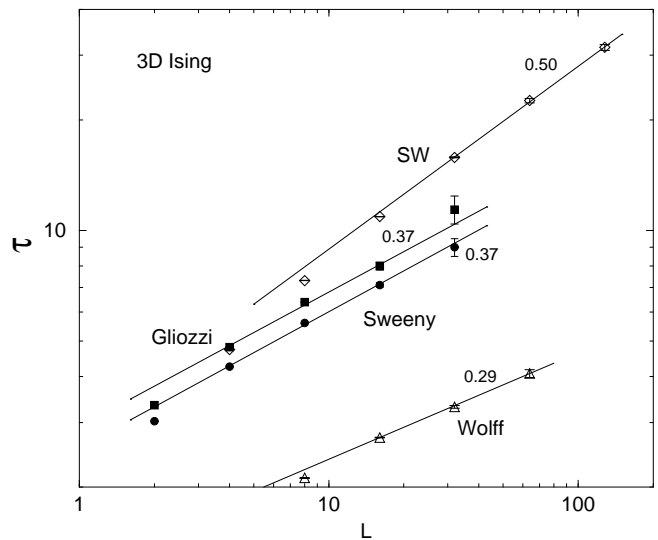}
\caption[3D Ising]{Correlation times $\protect\tau$ versus lattice linear 
dimension $L$ on a double logarithmic scale for the three-dimensional Ising 
model at $T_c$ of Sweeny (circles), Gliozzi (squares), Swendsen-Wang 
(diamonds), and Wolff (triangles) dynamics. The number on the straight line 
indicates the slope of the line.}
\label{fig-4}
\end{figure}

We performed our simulations at the critical temperature $kT_{c}(q)/J=1/\ln (%
\sqrt{q}+1)$ in two dimensions, and at $J/(kT_{c})=0.221657$ for the
three-dimensional Ising model \cite{landau}.  For each data point, we have
spent about two to four weeks of CPU time on 1GHz Pentium PCs.
For the Sweeny and Gliozzi dynamics, a unit of time $t$ for a lattice
linear dimension $L$ in $d$ dimensions is $d L^d$ moves with bond
selected at random.  We note that Gliozzi used a sequential sweep
of the bonds, which gives correlation time that is smaller by a factor
about 0.6. 
The correlation time data are given in Table~\ref{table1}.

Fig.~\ref{fig-2} contains a semilogarithmic plot of $\tau $ versus $L$ for
the four algorithms: Sweeny, Gliozzi, Swendsen-Wang, and Wolff. It is
remarkable that Gliozzi dynamics and to some extent Sweeny is perfectly
logarithmic in size down to $2\times 2$ lattice. For both Swendsen-Wang and
Wolff, some curvature can be seen in this plot, and the data can also be
fitted with a small exponent of 0.25 \cite{baillie} as well. We consider it
still to be an open question for Swendsen-Wang and Wolff dynamics whether
the correlation times depend on lattice sizes as a small power or
logarithmically \cite{heermann}.

In Fig.~\ref{fig-3}, results for the two-dimensional, three-state Potts
model are presented.  The straight lines in these log-log plots show good
power-law dependence for all algorithms. The data show an interesting
possibility that Sweeny and Gliozzi belong to a different dynamical
universality class from Swendsen-Wang. Linear least-squares fits give the
dynamical critical exponent $z=0.56\pm 0.01$ for Swendsen-Wang and $0.58\pm
0.01$ for Wolff, with lower values of $0.48\pm 0.01$ for Sweeny and $0.49\pm
0.01$ for Gliozzi dynamics. The differences are statistically significant.

Figure~\ref{fig-4} is for the three-dimensional Ising model. Again we see
similar behavior, although the data are less accurate comparing to 
that of two dimensions.  Sweeny and Gliozzi dynamics have approximately 
the same
dynamical critical exponent of $0.37\pm 0.02$, while that of the
Swendsen-Wang is larger at about $0.5$. 
Wolff single cluster algorithm gives 
remarkably small correlation times and a small exponent of
$0.29$ \cite{wolff3D}.  The Swendsen-Wang data show
some curvature and might approach the same asymptotic value for large
lattices. 

In summary, we have computed the correlation times for cluster algorithms.
It is clear that Sweeny and Gliozzi have the same dynamics. In all cases
studied here, the Sweeny rate is actually better than Gliozzi's. Both rates
reduce critical slowing down, but not completely eliminate it. Both Sweeny
and Gliozzi dynamics seem to have somewhat smaller dynamical critical
exponents than Swendsen-Wang when measured in units of Monte Carlo
sweeps, as we have done in this paper.

J.-S. W. acknowledges the hospitality of the Departments of Physics of
Carnegie Mellon University and Tokyo Metropolitan University, where part of
the work was done during a sabbatical leave. He also thanks Y. Okabe for
discussion.


\begin{thebibliography}{99}

\bibitem{gliozzi} F. Gliozzi, cond-mat/0201285.

\bibitem{sweeny} M. Sweeny, Phys. Rev. B \textbf{27}, 4445 (1983).

\bibitem{swendsen-wang} R. H. Swendsen and J.-S. Wang, Phys. Rev. Lett. 
\textbf{58}, 86 (1987).

\bibitem{wolff} U. Wolff, Phys. Rev. Lett. \textbf{62}, 361 (1989); Nucl.
Phys. \textbf{B322}, 759 (1989).

\bibitem{wu} F. Y. Wu, Rev. Mod. Phys. \textbf{54}, 235 (1982). 

\bibitem{FK} P. W. Kasteleyn and C. M. Fortuin, J. Phys. Soc. Jpn Suppl. 
\textbf{26}, 11 (1969); C. M. Fortuin and P. W. Kasteleyn, Physica \textbf{57%
}, 536 (1972).

\bibitem{chayes} L. Chayes and J. Machta, Physica A \textbf{254}, 477 (1998).

\bibitem{MK-Binder} H. M\"{u}ller-Krumbhaar and K. Binder, J. Stat.
Phys. \textbf{8}, 1 (1973).

\bibitem{landau-binder} See, e.g., D. P. Landau and K. Binder, \textsl{A
Guide to Monte Carlo Simulations in Statistical Physics}, pp. 91--93
(Cambridge University Press, 2000).

\bibitem{okabe} M. Kikuchi, N. Ito, and Y. Okabe, in \textsl{Computer
Simulation Studies in Condensed-Matter Physics VII}, p.44, Eds. D. P.
Landau, K. K. Mon, H.-B. Sch\"uttler, (Springer, Berlin, 1994).

\bibitem{baillie} C. F. Baillie and P. D. Coddington, Phys. Rev. B, \textbf{%
43}, 10617 (1991).

\bibitem{WangMCQMC} J.-S. Wang, cond-mat/0103318.

\bibitem{landau} A. M. Ferrenberg and D. P. Landau, Phys. Rev. B, \textbf{44}%
, 5081 (1991).

\bibitem{heermann} D. W. Heermann and A. N. Burkitt, Physica A \textbf{162},
210 (1990).

\bibitem{wolff3D} U. Wolff, Phys. Lett. B \textbf{228}, 379 (1989).

\end{thebibliography}
\end{document}